\documentstyle[11pt]{article}
\newcommand{\s}{\sigma }                                              
    
\newcommand{\tS}{\tilde{\Sigma}}
\newcommand{\tG}{\tilde G}    
\newcommand{\si}{\sigma _1}  
\newcommand{\st}{\sigma _2}

\newcommand{\xn}{x_{n}}

\newcommand{\xm}{x_{m}}

\newcommand{\e}{e^{i\int k_{0}Y}}

\newcommand{\kim}{ k_{1}^{\mu}}                                      
\newcommand{\kom}{ k_{0}^{\mu}}                                      
\newcommand{\ki}{ k_{1}}
\newcommand{\yn}{ Y_{n}}                                             
 
\newcommand{\kn}{ k_{n}}

\newcommand{\km}{ k_{m}}

\newcommand{\kt}{ k_{2}}                                             
\newcommand{\ko}{ k_{0}}                                             
\newcommand{\kob}{\bar{ k_{0}}}

\newcommand{\kin}{ k_{1}^{\nu}}                                      
\newcommand{\kon}{ k_{0}^{\nu}}                                      
\newcommand{\ktm}{ k_{2}^{\nu}}                                      
                                      
\newcommand{\lp }{\mbox {$e^{ia\int _{c} k(s) \partial _{z} X(z+as) ds
 + ik_{0} X(z)}$}}                                                    
    
\newcommand{\bp }{\mbox {$e^{i\int d t \int _{c} k(s, t )
\partial _{z} X(z+s, t )ds+ i\int dt \ko (t)X(t)}$}}

\newcommand{\dsix}{\frac{\partial}{\partial x_{2}(\sigma _{1})}}

\newcommand{\dst}{\frac{\partial }{\partial x_{2}}}

\newcommand{\dsii}{\frac{\partial ^{2}}
{\partial x_{1}^{2}}}
\newcommand{\p}{\partial}

\newcommand{\eps}{ \epsilon}                                        
\newcommand{\al}{\alpha }                                             
 
\newcommand{\tY}{\tilde Y}                                 
\newcommand{\lan}{\langle}
\newcommand{\ran}{\rangle}

\newcommand{\la}{\mbox{$ \lambda $}} 
\newcommand{\be}{\begin{equation}}
\newcommand{\br}{\begin{eqnarray}}
\newcommand{\ee}{\end{equation}} 
\newcommand{\er}{\end{eqnarray}}

\begin{document} 
\title{
\hfill\parbox{4cm}{\normalsize IMSC/2001/02/08\\
                               hep-th/0102017}\\        
\vspace{2cm}
 Wave Functionals, Gauge Invariant Equations for Massive Modes 
and the Born-Infeld Equation
 in the Loop Variable Approach to String Theory.
\author{B. Sathiapalan\\ {\em Institute of Mathematical Sciences}\\
{\em Taramani}\\{\em Chennai, India 600113}}}                                   \maketitle 

\begin{abstract} 
In earlier papers on the loop variable approach to gauge 
invariant interactions 
in string theory, a ``wave functional'' with some specific properties was 
invoked.
 It had the purpose of converting the generalized momenta to space time 
fields. 
In this paper we describe this object in detail and give some 
explicit examples. We also work out the interacting equations of the
massive mode of the bosonic string, interacting with electromagnetism,
and discuss in detail the gauge invariance. This is naturally described
in this approach as a massless spin two field interacting with a massless spin
one field in a higher dimension. Dimensional reduction gives the massive
system.
 We also show that in addition to describing fields 
perturbatively, as is required for reproducing the perturbative equations, 
the wave functional 
can be chosen to reproduce the Born-Infeld equations, which are 
non-perturbative
in field strengths. This makes contact with the sigma model approach. 

\end{abstract}
\newpage      
\setcounter{equation}{00}
\section{Introduction} 
    One of the outstanding issues in string theory is that of understanding
the fundamental symmetry principle underlying it. These symmetries
are partly manifested in the gauge invariance of the theory and also in the 
duality symmetries such as T-duality and S-duality. A formulation that
manifests some of these symmetries would be invaluable. The BRST formulation
of string field theory solves the technical problem of constructing a gauge
invariant action \cite{WS}-\cite{W1}. Nevertheless a space-time 
interpretation is obscured
by the heavy reliance on world sheet BRST properties. The sigma model approach
\cite{L}-\cite{BM}is somewhat more geometrical, 
but the issue of understanding gauge invariance
 or of going off-shell is rife with difficulty. Some progress on this has been
recently achieved \cite{KPP,BGKP}. 
    The loop variable approach motivated by the sigma model was 
used to obtain the 
free equations of motion in a simple manner \cite{BSLV}. 
It had the added advantage that the gauge transformations had a simple 
space-time interpretation of being scale transformations.  Speculations on the
symmetry principle of string theory are also contained in \cite{BSLV}.
   More recently, in papers \cite{BSLV1,BSLV2}
(hereafter I and II respectively)
 we had discussed a loop variable 
approach to gauge invariant string interactions.  In this paper we describe 
some concrete applications of this method.

The basic idea of the loop variable approach is to describe
the space time fields in string theory, using a generalized Fourier transform,
in terms of a ``wave functional'' $\Psi [k(s )]$. On mode 
expanding, the wave functional becomes a function 
$\Psi [\ki , \kt , ...\kn , ... ]$, where
$n$ runs over all natural numbers. 
 Thus 
\be
\Phi [X(z+s)] = \int {\cal D}k(s)e^{i\oint _c ds X(z+s) k(s)}\Psi[k(s)]
\ee

The free equations of motion are obtained by working with the
the loop variable $e^{i\oint _c ds X(z+s) k(s)}$. Actually the modified 
loop variable 
$\lp$ was used earlier and is more appropriate for our purposes \cite{BSLV}.
The gauge transformation is given by \cite{BSLV}
\be
k(s) \rightarrow k(s)\la (s) .
\ee

The interacting equation was obtained by broadening the loop to a band
whose width is parametrized by  $t$, so that $\ki$'s at different
$t$ correspond to different strings and at the same $t$ would correspond
to the same string. The loop variable thus takes the form:
\[
\bp
\]
and $\Psi [k(s,t)]$ is the form of the wave functional. 
Now the fully interacting gauge transformation is \cite{BSLV1,BSLV2}
\be
k(t,s)\rightarrow \int dt' \la (t',s)k(t,s).
\ee

An important feature is that this symmetry does not depend on world sheet
properties and holds even when there is a short distance cutoff on the
world sheet.

On mode expanding the wave functional qq
 becomes $\Psi [\ki (t), \kt (t),...\kn (t),...]$.
This is described in detail in I and II. The wave functional 
$\Psi$ was assumed to have certain 
properties defined below, but beyond that, was not discussed at all. In 
Sec 2 of this 
paper we discuss $\Psi$. We give an explicit expression for $\Psi$. A 
wave functional  
with these properties can describe the perturbative equations of motion.
In Section 3 and Section 4 we work out the equation for the first massive
mode of the string interacting with electromagnetism. This equation is valid
off the free mass shell also. The full equation has an infinite number of 
terms  involving all the massive modes. We keep only the massless and first
massive modes. We also show the gauge invariance to leading order 
in momenta and explain
the role of the higher modes in preserving gauge invariance in higher orders.

However if one wants to describe non perturbative equations with some specific
functional form for the fields, other forms of $\Psi$ are required. An example
is the Born-Infeld equation, describing uniform field strengths
\cite{CANY,FT2}. In Section 5 
we give a construction of $\Psi$ for this situation as well. 
In this approach, $z$, the position of the vertex operator, is also labelled by
$t$, i.e. $z(t)$. Since $z$ is real for an open string, it is possible to
let $z(t)=t$. This is done in this paper and one makes more direct contact
with the open string sigma-model or boundary string field theory
approaches \cite{FT2,CANY,BSPT,W2,WL,SS}. 

\setcounter{equation}{0}
\section{$\Psi$ for Perturbation Theory} 
Let us recollect the basic properties required of $\Psi$.
\[
 \int [\prod _{n > 0}{\cal D} \kn (t )]\Psi [\ko (t), \kn (t)]= 
\prod _t \delta (\ko (t)) \]

In position space this equation would read:
\[
\int [\prod _{n > 0}{\cal D} \kn (t )]\Psi [X (t), \kn (t)]=
\int [\prod _{n\ge 0}{\cal D} \kn (t )]e^{i\int dt \ko (t) X (t)}
\Psi [\ko (t), \kn (t)]=1
\]

\[
 \int [\prod _{n > 0}{\cal D} \kn (t )]
 \kim (t_1 )\Psi [\ko (t ) ,\kn (t ) ] = A^\mu (\ko (t_1 ))
\prod _{t\ne t_1} \delta (\ko (t)) 
\]
In position space:
\[ 
\int [\prod _{n > 0}{\cal D} \kn (t )]
\kim (t_1 )\Psi [X (t ) ,\kn (t ) ] = A^\mu (X(t_1 )) ,
\]

\[
\int [\prod _{n > 0}{\cal D} \kn (t )]
\kim (t_1 )\kin (t_2 )\Psi [\ko (t ) ,\kn (t ) ] = A^\mu (\ko (t_1 ))
A^\nu (\ko (t_2 ))\prod _{t\ne t_1,t_2} \delta (\ko (t))
\]
\be \label{2.02}
+ \delta (t_1-t_2)S_{1,1}^{\mu \nu }(\ko (t_1 ))\prod _{t\ne t_1}
 \delta (\ko (t)) .
\ee

In position space:

\[
\int [\prod _{n > 0}{\cal D} \kn (t )]
\kim (t_1 )\kin (t_2 )\Psi [X (t) ,\kn (t ) ] = A^\mu (X (t_1 ))
A^\nu (X (t_2 ))
\]
\be \label{2.03}
+ \delta (t_1 - t_2 )S_{1,1}^{\mu \nu }(X (t_1 )) .
\ee

A word on the notation used in these equations: $\ko (t)$ refers to the
momentum of the vertex operator located at the point $z(t)$($=t$ in this
 paper).  $t$ is only a label. Thus while $X(t)$ is a non-trivial function 
of $t$, $\ko (t)$ is not. Further in the end all momenta are integrated over.
Thus $\ko (t)$ is only an integration variable: $\int d\ko (t)\phi (\ko (t))
e^{i\ko (t)X}=\phi (X)$.  It also follows that $\phi (\ko (t))$ is not 
explicitly or implicitly a function of $t$.

We can easily construct a wave functional, $\Psi$, with these properties.
Let us first set  $S_{1,1}^{\mu \nu}=0$. Then the solution is very simple:
\be  \label{2.04}
\Psi [X(t),\kn (t)]= \prod _{t,\mu} \delta [\kim (t)-A^\mu (X(t))]
\prod _{t,n> 0,\mu}\delta [\kn ^\mu (t)]
\ee

 In momentum space one can write
\be  \label{2.05}
\Psi [\ko (t), \kn (t)]= \int {\cal D}X(t) e^{-i\int dt \ko (t) X(t)}
\prod _{t,\mu}
\delta [\kim (t)-A^\mu (X(t))]\prod _{t,n> 0,\mu}\delta [\kn ^\mu (t)]
\ee
The generalization, to include the massive modes, $S_n^\mu ,n>1$ is obvious.

One can easily check that $\Psi$ satisfies (\ref{2.03}). To make sense of the
 product over $t$, one has to regularise. This is done below.

Let us introduce $S_{1,1}^{\mu \nu}$.
This can be done by broadening the delta function to include a non-trivial
two-point function. Let us first discretise the parameter $t$  and set
$t=na$ where $n \in {\bf Z}$ and $a$ is a regulator. Then the wave functional

\be
\Psi [\ki (t),\km (t), X(t)] = \lim _{N\rightarrow \infty}
   [ \prod _{n=-N/2}^{N/2}
{\prod _{m> 0}\delta [\km (n)]\over \sqrt {2\pi S_{1,1}(X(n))\over a}}]
e^{-\sum _{n=-N/2}^{N/2} {[\ki (n)- A_1(X(n))]^2a\over 2S_{1,1}(X(n))}}
\ee
  This satisfies
\[
\langle \ki (n_1) \rangle = A(X(n_1))
\]
\be
\langle \ki (n_1)\ki (n_2)\rangle - \langle \ki (n_1)\rangle 
\langle \ki (n_2)\rangle = S_{1,1}(X(n_1)){\delta _{n_1,n_2}\over a}
\ee
where $\langle ... \rangle = \int [\prod _{n> 0}{\cal D} \kn ]...\Psi [\kn (t),X(t)]$.

We can Fourier transform $X(n)$ and write down the momentum space wave 
functional
\[
\Psi [\ki (t),\km (t), \ko (t)] = 
\]
\be   \label{wfk}
\lim _{N\rightarrow \infty}
   [ \prod _{n=-N/2}^{N/2} dX (n)e^{-ia\ko (n) X(n)}
{\prod _{m> 0}\delta [\km (n)]\over \sqrt {2\pi S_{1,1}(X(n))\over a}}]
e^{-\sum _{n=-N/2}^{N/2} {[\ki (n)- A_1(X(n))]^2a\over 2S_{1,1}(X(n))}}
\ee

From (\ref{wfk}) it is clear that the momentum variable conjugate to $X(n)$
is $a\ko (n)$ and thus in the inverse Fourier transform, the measure
 ${\cal D}\ko (t)$
is to be interpreted as $\prod _n (a\, d\ko (n))$. Let us call
$a\ko = \kob$

If we define in the usual way
\[
\int dX e^{-ikX} A(X) = A(k)
\]
Then 
\[
\int dX e^{-iakX} A(X) = A(ak)= A(\kob )
\]

The wave functional (\ref{wfk}) satisfies

\[
\langle \kim (n_1) \kon (n_2)\rangle =
          A^\mu (\kob (n_1))\kob ^\nu (n_1){\delta _{n_1,n_2}\over a}
\prod _{n\ne n_1}
\delta [\kob (n)].
\]
Here $\kob$ is the (physical) momentum conjugate to the position. Since
the measure of integration is also $d\kob$ this overall factor of $a$
is not important.

Note also that in momentum space,
\[
\langle 1 \rangle = \prod _n \delta [\kob (n)],
\]
This is the regularized version of the first equation in (\ref{2.02}),
which is the statement that the vacuum is translationally invariant.
\footnote{We have set the tachyon field to zero in this paper}

We can also include the gauge transformations of the loop variable by 
incorporating $\la _1(t), \, \la _2 (t)...$into the wave functional.
  Let us consider (\ref{2.04}).

\[
\Psi [X(t),\kn (t)]= \prod _{t,\mu} \delta [\kim (t)-A^\mu (X(t))]
\prod _{t,n> 0,\mu}\delta [\kn ^\mu (t)]
\]

One can incorporate $\la _1 (t)$ by adding a delta function
\be  \label{ALa}
\Psi [X(t),\kn (t),\la _1(t)]= \prod _{t,\mu} \delta [\kim (t)-A^\mu (X(t))]
\prod _{t}\delta [\la _1(t)-\Lambda (X(t))]
\ee

In momentum space
\[
\Psi [\ko (t),\kn (t),\la _1(t)]=
\int {\cal D}X(t)e^{-i\int dt \ko (t) X(t)}\Psi [X(t),\kn (t),\la _1(t)]
\]
Regularization is understood and has not been explicitly shown.

The above wave functional satisfies
\[
\langle \la _1(n_1)\rangle = \Lambda (\kob (n_1))
\]

\[
\langle \la _1 (n_1)\kom (n_2)\rangle = \Lambda (\kob (n_1) )\kob  ^\mu (n_2)
{\delta _{n_1,n_2}\over a}\prod _{n\ne n_1}
\delta [\kob (n)].
\]
as required. $\la _n$ , $n>2$, can also similarly be included.

\setcounter{equation}{00}
\section{Gauge Transformations of Space-Time Fields}

In this section we show that gauge transformations can be consistently
defined for space-time fields. 

In II, a scheme was described for defining gauge transformations
on space-time fields.  The basic idea was to first move all vertex operators
to one location and then use these vertex operators to define combinations
of fields (and loop variables) whose gauge transformations are then worked out.

Thus
\[
e^{i\sum _n\kn (z_1)\tY 
_n (z_1)}=
e^{i\sum _nK_n (z_1, z-z_1)\tY _n (z)}
\]
This equation defines the $K_n$. The precise expressions for $K_n(z_1,z-z_1)$
are given in II.
All the $z_i$ dependence is thus made 
explicit. Then we consider combinations of the form
\[
\prod_{i=1}^N K_{n_i}(z_i,z-z_i)\tY _{n_i}(z).
\]
This is used to define the transformation laws for the highest level
field $\lan k_{n_1}k_{n_2}...k_{n_N}\ran = S_{n_1,n_2,...n_N}$ in terms
of the gauge transformation of lower fields. In this way gauge transformation 
laws for all fields can be recursively defined. 

     The crucial point in this construction is that we 
are using a regularized greens function that has no short distance singularity.
This is what allows us to move all vertex operators to one point, and also 
allows us to contract them if necessary. Furthermore we will also integrate
the location of the vertex operator over some (finite) range - this is the 
usual Koba-Nielsen integration. In the examples considered in \cite{BSLV2}
the $z$-dependence somehow dropped out and it wasn't necessary to do the 
integrals.
In general, the $z$-dependence will not drop out (indeed, they are essential)
 and only the integrated equation makes sense.
The combination of operators used here for defining the gauge transformations
are precisely the combinations that occur in any equation of motion - along 
with contractions of the $\tY$'s.  Since this contraction does not affect
the consistency of this procedure it follows that there is a well defined 
map from loop variables and their gauge transformation to space-time
fields and their gauge transformation.  Thus gauge invariance at the level
of loop variables implies gauge invariance of the field equations.

Below we discuss the gauge invariance of the $Y_2$ equation of \cite{BSLV1}.
We will demonstrate the leading order gauge invariance explicitly. For the 
non-leading term we will show how massive modes contribute in a crucial way
by providing the types of terms required for gauge invariance.
Explicit calculations of numerical coefficients is rather tedious and will not
be attempted.

We will include the contributions of some of the massive modes and 
will go some way in proving the gauge invariance of that equation - 
far enough to make the pattern clear. 

We use $ln \, (\eps ^2 + (z_1-z_2)^2)$ as the two point function $\langle 
X(z_1) X(z_2)\rangle $. If we use this, the equation of motion corresponding 
to $Y_2$ from \cite{BSLV1} (combining equations (5.3.39),(5.3.40), (5.3.45), 
(5.3.50) , (5.3.51) and (5.3.54)) becomes the sum of various terms listed below:
\be  \label{E.1}
- A(p).(p+q)iA^\mu (q) |{(z-w)^2\over \epsilon ^2}+1|^{p.q}(\epsilon )^
{(p+q)^2}Y_2^\mu e^{i(p+q)Y}
\ee
\be     \label{E.2}
-S_{1,1}^{\mu \nu}\kon iY_2^{\mu}e^{i\ko Y} (\epsilon )^{\ko ^2}
\ee
\be     \label{E.3}
- (\epsilon )^{\ko ^2}S_2(\ko ).\ko i\kom Y_2 ^\mu e^{i\ko Y}
\ee
\be   \label{E.4}
(\epsilon )^{(p+q)^2}|{(z-w)^2\over \epsilon ^2 }+1 |^{p.q}A(p).A(q)i(p+q)^\mu
Y_2^\mu
e^{i\ko Y}
\ee
\be    \label{E.5}
(\epsilon ) ^{\ko ^2} S^\nu _{1,1\nu} (\ko )i\kom Y_2 ^\mu e^{i\ko Y}
\ee

\be \label{E.6}
(\epsilon )^{\ko ^2} \ko ^2 iS_2^\mu (\ko ) Y_2^\mu e^{i\ko Y}
\ee

The notation is that of I and II. The reader need only note that after 
obtaining the equations of motion one can set $\xn =0$ (again in the notation
of I and II) and after that $Y_2 \equiv \tY _2 \equiv \p _z^2 X$ and 
$Y \equiv X$.

The gauge transformations of the fields are (according to the method defined
above and in \cite{BSLV2}) :
 
\br
\delta S^{\mu \nu}_{1,1} (k)e^{i\ko Y}&=& [\Lambda _{1,1}^{(\mu} (k) 
\ko ^{\nu )} + \delta _{int} S^{\mu \nu}_{1,1}(k)]e^{i\ko Y} \nonumber \\
\delta A_1^\mu (p)e^{i p_0 Y} & = &p^\mu \Lambda _1(p)e^{ip_0 Y} 
\er

 and
\be     \label{dS}
\delta _{int}S_{1,1}^{\mu \nu}(k) =
\int dp dq \delta (p+q-k)[\Lambda _1(p)q^{(\nu} A_1^{\mu )} (q)]
\ee

\be
\delta S_2^\mu (\ko )= \Lambda _2 (\ko )\kom + \Lambda _{1,1}^\mu (\ko
) + \delta _{int}S_2^\mu (\ko )
\ee

with 

\be
\delta _{int}S_2^\mu (\ko ) =  \int dp dq \delta (p+q-\ko )\Lambda
_1(p)A_1^\nu (q)
\ee

Let us write down the contribution of the various terms:

\br
& \delta (\ref{E.1})=  \nonumber \\
& (-p.(p+q)\Lambda (p)iA^\mu (q) - A(p).(p+q) iq^\mu \Lambda
(q) ) [1+ {(z-w)^2\over \epsilon ^2}]^{p.q}\epsilon ^{(p+q)^2}\nonumber \\
&   
\er

\br
& \delta (\ref{E.2})= \nonumber \\ & -\Lambda _{1,1}^{(\mu}k_0 ^{\nu )}\kon
i\eps ^{\ko ^2} - \Lambda _1(p)q^{(\nu }A_1^{\mu )}(q)(p+q)^\nu i \eps ^{(p+q)^2}
\nonumber \\
&   
\er

\br
& \delta (\ref{E.3}) = \nonumber \\ 
& -\Lambda _2 (\ko )\ko ^2 i\kom \eps ^{\ko ^2}  
-\Lambda _{1,1}(\ko ).\ko i\kom \eps ^{\ko ^2}
-\Lambda _1 (p)A_1(q).(p+q)
i(p+q)^\mu \eps ^{(p+q)^2}
\nonumber \\ &   
\er

\br
& \delta (\ref{E.4}) = \nonumber \\
& 2p.A(q) \Lambda (p) i(p+q)^\mu  [1+ {(z-w)^2\over \epsilon ^2}]^{p.q}
\eps ^{(p+q)^2}
\er

\br
& \delta (\ref{E.5}) = \nonumber \\ & 2\Lambda _{1,1}(\ko ).\ko i\kom 
\eps ^{\ko ^2}+ 
2\Lambda _1 (p)q.A_1(q)i(p+q)^\mu \eps ^{(p+q)^2}\nonumber \\
&   
\er

\br
& \delta (\ref{E.6}) = \nonumber \\ &
\ko^2 i\Lambda _2 (\ko )\kom \eps ^{\ko ^2} + i\Lambda ^\mu _{1,1}(\ko )\ko ^2
\eps ^{\ko ^2} +i\Lambda (p) A_1^\mu (q)(p+q)^2 \eps ^{(p+q)^2} \nonumber \\&  
\er

In the above expressions, integration over momenta $p,q$ with momentum 
conserving delta function, and the Koba-Nielsen 
variables $w,z$ with suitable regularization, are understood.

 Terms involving $\Lambda _2$ and $\Lambda _{1,1}$ can easily be seen to add up
to zero. We are left with
\[
A(p).(p+q) \Lambda _1(q)[-iq^\mu  [1+ {(z-w)^2\over \epsilon ^2}]^{p.q}
 -ip^\mu 
-i(p+q)^\mu] 
\]
\be
 + A(p).p \Lambda _1(q)2i(p+q)^\mu + A(p).q\Lambda _1(q) 2i(p+q)^\mu 
 [1+ {(z-w)^2\over \epsilon ^2}]^{p.q}
\ee
and
\be
iA^\mu (q)\Lambda _1(p)[-p.(p+q) [1+ {(z-w)^2\over \epsilon ^2}]^{p.q}
-q.(p+q) + (p+q)^2]
\ee

If one expands  $[1+ {(z-w)^2\over \epsilon ^2}]^{p.q} 
\approx 1+p.q{(z-w)^2\over \epsilon ^2} + ... O((z-w)^4) $ in the above 
expressions, 
it is easy to see that the leading terms cancel leaving uncancelled 
terms of order $(z-w)^2$. 

Where does one find contributions to cancel these terms? The answer is that
 they come from the variations of higher modes such as 
$S_{1,1,2}^{\mu \nu \rho}$. These higher modes do contribute to the $Y_2$
 equation through the following terms in the loop variable

\[
\int \int \int \int dt_1 dt_2 dt_3 dt_4
 \ki (t_1).\ki (t_2)
{\frac{\partial ^{2}[\tS +\tilde G] (t_1 ,t_2 )}
{\partial
 x_{1}(t_1 )\partial x _{1}(t_2 )}}
\]
\be  \label{E.8}
\kt (t_3).\ko (t_4)
{\frac{\partial [\tS +\tilde G] (t_3 ,t_4 )}
{\partial
 x_{2}(t_3 )}}e^{\int dt \int dt' \ko (t).\ko (t') [\tS + \tG ](t,t')}\e
\ee

and also 
\[
\int \int dt_1 dt_2 \ki (t_1).\ki (t_2)
{\frac{\partial ^{2}[\tS +\tilde G] (t_1 ,t_2 )}
{\partial
 x_{1}(t_1 )\partial x _{1}(t_2 )}}
\]
\be  \label{E.9}
\int dt_3
e^{\int dt \int dt' \ko (t).\ko (t') [\tS + \tG ](t,t')}
i\ktm (t_3) Y_2^\mu (t_3)\e
\ee

using $ {\frac{\partial ^{2}\tS (t_1 ,t_2 )}
{\partial
 x_{1}(t_1 )\partial x _{1}(t_2 )}}\mid _{\xn (t_1)=\xn (t_2)}
\approx {1\over 2}(\dsii - \dst )\Sigma$ 
,$\dsix \tS \mid _{\xn (t_1)=\xn (t_2)}
\approx  {1\over 2}\dst \Sigma$, \,  $\dst \tG \mid _{t_1=t_2,\xn =0}=
 {1\over \eps ^2}$, and \, $  {\frac{\partial ^{2}\tG (t_1 ,t_2 )}
{\partial
 x_{1}(t_1 )\partial x _{1}(t_2 )}}\mid _{\xn =0)}= 
-{1\over \eps ^2}$ we get on varying $\Sigma$ in
(\ref{E.8}) and taking expectation of the loop variable,
\be
{3\over 2\eps ^2}\langle \ki .\ki \kt .\ko i\kom \rangle Y_2^\mu 
\approx {3\over 2\eps ^2} S_{1,1,2}^{\rho \rho \nu}(\ko )\kon i\kom Y_2^\mu
\ee 
 where we have kept only the $S_{112}$ ``contact'' term in the expectation 
value.

The second term, (\ref{E.9}), gives analogously
\be
-{1\over \eps ^2} \langle \ki .\ki \ko ^2 i\ktm \rangle Y_2^\mu
\approx -{1\over \eps ^2}\ko ^2 S_{1,1,2}^{ \rho  \rho   \mu}(\ko )Y_2^\mu
\ee

There are also terms of the form $\ki .\kt \ki .\ko ,\, \ki .\kt \kim $ 
that can contribute to the $Y_2$ equation. Since 
we are only interested in showing
that these terms contribute $O(z^2)$ terms to the equation of motion, 
and not in the precise coefficient, we will not worry about it here.

Let us now turn to the scheme of \cite{BSLV2} to evaluate 
$\delta S_{1,1,2}^{\mu \nu \rho}$.

Let us start with $K^\mu _1 (z_1,z-z_1)K^\nu _1(z_2, z-z_2)
\tY _1^\mu (z)\tY _1^\nu (z)$.

\[
\int dz_1 \int dz_2
\langle K^\mu _1 (z_1,z-z_1)K^\nu _1(z_2, z-z_2) \rangle
\]
\[
=
\int dz_1 \int dz_2 
\langle (\kim (z_1) +(z_1-z)\kom ) (\kin (z_2) +(z_2-z)\kon )\rangle 
\]
\[=
\int dz_1 \int dz_2 
S_{1,1}^{\mu \nu}(\ko (z_1))\delta (z_1-z_2)+
 A_1^\mu (\ko (z_1))A_1^\nu (\ko (z_2 ))
\]
\[+ (z_1-z)\kom (z_1)A_1^\nu (\ko (z_2))\delta (z_1-z_2)+
(z_2-z)\kon (z_2)A_1^\mu (\ko (z_1))\delta (z_1-z_2)
\]
\[=
\int dz_1 S_{1,1}^{\mu \nu }(\ko (z_1))+ \int dz_1 \int dz_2 
A_1^\mu (\ko (z_1))A_1^\nu (\ko (z_2 ))
\]
\be +
\int dz_1 (z_1-z)\ko ^{(\mu }(z_1)A_1^{\nu )}(\ko (z_1))
\ee

We remind the reader about the following
two points. First, $\ko (z)$ refers to a momentum of a 
field associated with a particular vertex operator located at $z$.
 $\ko$ is {\em not} a function of $z$. In the same way $S_{1,1}(\ko (z))$ is a 
field with momentum $\ko (z)$ - it is not a function, implicitly or
 explicitly, of $z$. The $z$-dependence in this approach is all explicitly
there in the equations because we have extracted all the $z$-dependences
by Taylor expanding the vertex operators. Second, integration of $z_i$
over some suitable range is understood in all these equations.\footnote{Thus
$\int dz_1 \equiv \int _a^R dz_1 $ for some $a,R$ that we will leave
unspecified. If we are working on a unit disk then $z_1$ has a range of 
$2\pi$.  If on the upper half plane, $z_1$ ranges from $-\infty$ to $+\infty$.
In the context of the proper time formalism, $z$ would range from $0$ to $R$.
This scheme dependence is expected in any off shell description.}  

We now consider the variation of both sides.

\[
\delta \int dz_1 \int dz_2 
\langle (\kim (z_1) +(z_1-z)\kom ) (\kin (z_2) +(z_2-z)\kon )\rangle 
\]
\be
=
\int dz' \int dz_1 \int dz_2 
\langle \la _1(z')(\kom (z_1) (\kin (z_2) +(z_2-z)\kon )
+ (\kim (z_1) +(z_1-z)\kom ) \kon (z_2)  )\rangle 
\ee
Thus variation of the space-time fields gives:
\[
\delta \{
\int dz_1 S_{1,1}^{\mu \nu }(\ko (z_1))+ \int dz_1 \int dz_2 
A_1^\mu (\ko (z_1))A_1^\nu (\ko (z_2 ))
\]
\[+
\int dz_1 (z_1-z)\ko ^{(\mu }(z_1)A_1^{\nu )}(\ko (z_1))\}
\]
\[
=
\int dz_1 [\delta S_{1,1}^{\mu \nu}(\ko (z_1))] +
\int dz_1 \int dz_2 [\Lambda _1 (\ko (z_1))\ko ^{(\mu}(z_1)A_1^{\nu )}
(\ko (z_2))]
\]
\be + \int dz_1
 [(z_1-z)\ko ^{(\mu}(z_1)\Lambda
_1 (\ko (z_1))\ko ^{\nu )}(z_1)].
\ee
Whereas variation of the loop variable momenta yields

\[
\int dz' \Lambda _{1,1}^{(\nu}(\ko (z'))\ko ^{\mu )}(z') + \int dz'dz''
\Lambda _1(\ko (z'))A_1^{(\nu}(\ko (z'))[\ko (z') + \ko (z'')]^{\mu )}
\]
\be +
\int dz' 2\Lambda _1 (\ko (z'))\kom (z')\kon (z')(z'-z)
\ee

Comparison of the two equations gives:

\[
\int dz_1 \delta S_{1,1}^{\mu \nu}(\ko (z_1))= 
\]
\be  \label{S11}
\int dz' \Lambda _{1,1}^{(\nu}
(\ko (z'))\ko ^{\mu )} (z') + \int dz_1 dz_2 \Lambda _1 (\ko (z_1))A_1^{(\nu}
(\ko (z_2))\ko ^{\mu )}(z_2)
\ee

One can extract from this $\delta S_{1,1}^{\mu \nu }(p)$ if one uses the fact that the integral over $z'$ in the LHS of (\ref{S11}) gives simply
$range \, of \, z' \times \delta S_{1,1}^{\mu \nu }$.

As pointed out in \cite{BSLV2} earlier, there are no $z$-dependent terms 
in this variation.
However when we turn to $S_{1,1,2}$ this will not be true.

Let us turn to $S_{1,1,2}$:

We start with
\[
\int dz_1 \int dz_2 \int dz_3
\langle K_1^\mu (z_1,z-z_1) K_1^\nu (z_2,z-z_2)K_2^\rho (z_3,z-z_3) \rangle
\]
\[
=
\langle (\kim (z_1) + (z_1-z)\kom (z_1))(\kin (z_2)+(z_2-z)\kon (z_2))
(\kt ^\rho (z_3)+(z_3-z)\ki ^\rho + {(z_3-z)^2\over 2} \ko ^{\rho})\rangle
\]

We will focus our attention on the $O(z^2)$ terms, which are:

\[
\langle \kim (z_1)\kin (z_2)\ko ^\rho (z_3){(z_3-z)^2\over 2}+
\kom (z_1)\kon (z_2)\kt ^\rho (z_3)(z_1-z)(z_2-z)+
\]\be \label{k112}
\kom (z_1)\kin (z_2)\ki ^\rho (z_3)(z_1-z)(z_3-z)+
\kim (z_1)\kon (z_2)\ki ^\rho (z_3)(z_3-z)(z_2-z)\rangle
\ee

Taking the expectation value one gets (Integration over $z_i$ is understood):
\[
S_{1,1}^{\mu \nu }(\ko (z_1))\ko ^\rho (z_1){(z_1-z)^2\over 2}+
\]\[
A_1^\mu (\ko (z_1))A_1^\nu (\ko (z_2))[\ko ^\rho (z_1){(z_1-z)^2\over 2}+
\ko ^\rho (z_2){(z_2-z)^2\over 2}]+
\]
\[
\{ S_{1,1}^{ \nu \rho }(\ko (z_1))\kom (z_1)(z_1-z)^2 +\]\[
A_1^\nu (\ko (z_2))A_1^\rho (\ko (z_3))[\kom (z_2)(z_2-z)+\kom (z_3)(z_3-z)]
(z_3-z)+
\]
\[
+ \mu \leftrightarrow \nu \} \, +
\]
\be
S_2^\rho (z_2)\kom (z_3) \kon (z_3)(z_3-z)^2
\ee

For clarity we will rewrite this expression with momenta $p.q$.
This gives

\[
S_{1,1}^{\mu \nu }(p)p ^\rho{(z_1-z)^2\over 2}+
\]\[
A_1^\mu (p)A_1^\nu (q)[p ^\rho {(z_1-z)^2\over 2}+
q ^\rho {(z_2-z)^2\over 2}]+
\]
\[
\{ S_{1,1}^{ \nu \rho }(p)p^\mu(z_1-z)^2 +\]\[
A_1^\nu (q)A_1^\rho (p)[q^\mu(z_2-z)+p^\mu (z_3-z)]
(z_3-z)+
\]
\[
+ \mu \leftrightarrow \nu \} \, +
\]
\be   \label{S112}
S_2^\rho (p)p^\mu p^\nu (z_3-z)^2
\ee

We have to compare the gauge transformation of this expression (using
the previously determined variations of the fields), and compare with what
 one gets on varying the loop variable expression directly. If they are not equal,
the difference will be assigned to the gauge transformation of $S_{1,1,2}$.
Let us focus on terms having the structure 
$A_1^\nu (q)\Lambda _1(p)p^\mu q^\rho$.

Note that in (\ref{S112}) there are two contributions that have this structure:
One is multiplied by $(z_3-z)^2$ and the other by $(z_2-z)^2$, which is the 
same
since $z_i$ are integration variables.

Now we vary the $k_i$ in (\ref{k112}) and look at the terms that can give
$A_i^\nu$. These  are:
\[
\int dz' \langle \la _1 (z')(\kom (z_1)\kin (z_2) 
\ko ^\rho (z_3){(z_3-z)^2\over 2} +
\]\[
\kom (z_1) \kin (z_2) \ko ^\rho (z_3)(z_1-z)(z_3-z))\rangle
\]
The second term gives a contribution proportional to $\Lambda _1 (p)p^\mu 
A_1^\nu (q) q^\rho (z_1-z)(z_2-z)$.
which is clearly different from anything obtained from (\ref{S112}).
 This term has thus to be assigned to $\delta
S_{1,1,2}^{\mu \nu \rho}$. Thus we have shown that there are $O(z^2)$ terms
in the gauge variation of $S_{1,1,2}$ contribution to the $Y_2$ equation of
 motion.  

The above calculation gives us a picture of how the contributions 
from the higher modes to a given equation of motion add up to reproduce
expressions of the form $\int dz \int dw (z-w)^{2p.q}$. As mentioned in
\cite {BSLV2} , the fact that the equations are gauge invariant follows
from the gauge invariance of the loop variable expression and the fact 
that the above map - from loop variables to space-time fields and their 
respective gauge transformations  - is well defined.
  Furthermore
the space-time gauge invariance does not depend on any world sheet properties
such as BRST invariance, so this is possible. This also allows one to go
off-shell without violating gauge invariance.

\section{Dimensional Reduction}
\setcounter{equation}{00}

We can dimensionally reduce the above equations to obtain equations for 
the massive modes with masses equal to that in string theory. This is where
we make contact with string theory. The prescription was given in 
\cite{BSLV2}. We start with the loop variable:   
\[ 
exp \{
\int d\si \int d\st 
\sum _{n,m\ge 0}(\kn (\si ) .\km (\st ) {\p ^2 [\tG +\tS ] (\si ,\st )
 \over \p \xn (\si )\p \xm (\st )}
+ 
\]
\be   \label{newLV}
 k_{n, V} (\si )k_{m,V} (\st ) 
{\p ^2\tS  (\si ,\st ) \over \p \xn (\si )\p \xm (\st )}) \}
:exp \{ i\int d\s \sum _{n\ge 0}\kn \yn (z(\s )\}:.
\ee

The extra dimension (27th in the bosonic string, or 11th in the superstring)
is denoted by $V$ in the above equation. 
In \cite{BSLV2} we set $\ko ^V ={\sqrt{P-1}\over N}$ where
$N$ is the number of fields in a particular term in the equation of motion.
(Thus $N=1$ for the free theory, $N=2$ in the quadratic term of an 
equation of motion, etc.) and $P$ is the engineering dimension
 of the vertex operator
for which the equation is being written. Thus $P=0$ for $e^{ikX}$, $P=1$ for
$\p _zX e^{ikX}$ ,  etc. Note that if some of the $Y$'s are contracted
then we get powers of $\eps$, and $P$ must count these powers also as 
contributing to the engineering dimension.  
 Furthermore we will let
\[
\langle \kn ^V \rangle =  S_n^V
\]

The factor $1\over N$ in 
$\ko ^V$ can also be thought of as being equivalent to the imposition
 of momentum conservation
in the extra dimension
so that the total momentum of each term in an equation adds up to the same 
value as for the linear term, where it is set equal to the mass of the field.

Let us apply this to  the first massive level equation of the last section.
This has $P=2$. 

First we have to establish the gauge transformation laws for the fields
$A_1^V , S_{1,1}^{VV}, S_{1,1}^{\mu V}$, and $S_2^V$. We start with
\[
\langle K_1^V(z_1,z-z_1) \rangle Y_1^V = 
\langle (\ki ^V (z_1) + (z_1-z)\ko ^V )\rangle Y^V_1
\]
\be
= A_1^V (\ko (z_1))Y^V_1
\ee
\be  \label{dav}
\delta \langle K_1^V(z_1,z-z_1) \rangle Y^V_1 = 
\int dz' \langle \la _1 (z') \ko ^V(z_1) \rangle Y^V_1
\ee
Since $P=1$, $\ko ^V =0$ in the above equation.
So,
\be   \label{AV}
\delta A_1^V(\ko ) = 0
\ee

This is just as well, since for making contact with string theory, we will
set this field to zero. Note also, that if $Y^V_1$ in (\ref{dav}) is 
contracted with $Y^V$ in $e^{i\ko ^VY^V}$ we get
(Using $\lan Y_1^V (z)Y^V(z)\ran = \eps _1 \approx {1\over \eps}$)
\[
\delta \langle \int dz K_1^V(z_1,z-z_1)\ko ^V (z)\rangle \eps _1 = 
\int dz' \langle \int dz \la _1 (z') \ko ^V(z_1)\ko ^V (z) \rangle \eps _1 =0
\]
since $(P-1)\eps _1 =0$. Thus the above procedure is consisitent with
contractions.

 We now turn to
\[
\int dz_1 \int dz_2
\langle K^V _1 (z_1,z-z_1)K^V _1(z_2, z-z_2) \rangle Y^V_1Y^V_1
\]
\[
=
\int dz_1 \int dz_2 
\langle (\ki ^V (z_1) +(z_1-z)\ko ^V ) (\ki ^V (z_2) +(z_2-z)\ko ^V )\rangle 
Y^V_1Y^V_1
\]
\[
=[\int dz_1 \int dz_2 
S_{1,1}^{V V}(\ko (z_1))\delta (z_1-z_2)+
 A_1^V (\ko (z_1))A_1^V (\ko (z_2 ))
\]
\[+ (z_1-z)\ko ^V (z_1)A_1^V (\ko (z_2))\delta (z_1-z_2)+
(z_2-z)\ko ^V (z_2)A_1^V (\ko (z_1))\delta (z_1-z_2)]Y^V_1Y^V_1
\]
Since $A_1^V=0$ we have
\be=
\int dz_1 S_{1,1}^{V V }(\ko (z_1))Y^V_1Y^V_1
\ee

We turn to the variations.

\[
 \delta \lan K_1^V K_1^V \ran Y^V_1Y^V_1
= \int dz' \int dz_1 \int dz_2 \lan \la (z') (\ko ^V (z_1) \ki ^V (z_2) + 
\ki ^V (z_1) \ko ^V (z_2) + 2\ko ^V (z_1) \ko ^V (z_2) (z_2-z))\ran 
Y^V_1Y^V_1
\]
\[
= 2\int dz_2 \Lambda _{1,1}^V (z_2)Y^V_1Y^V_1
+ 2\int dz' \underbrace {\ko ^V (z')}_{1} \underbrace {\ko ^V (z')}_{1}
 (z'-z)\Lambda _1 (z')Y^V_1Y^V_1
\]
\be   \label{2.05}
=
2\int dz_2 [\Lambda _{1,1}^V (z_2) +(z_2-z)\Lambda _1 (z_2)]Y^V_1Y^V_1
\ee

Again we have set $A_1^V=0$.

Thus we get

\be  \label{SVV}
\delta \int dz _2S_{1,1}^{VV} = 2\int dz_2 
[\Lambda _{1,1}^V (z_2) +(z_2-z)\Lambda _1 (z_2)] 
\ee

We now turn to

\[
\int dz_1 \int dz_2
\langle K^V _1 (z_1,z-z_1)K^\mu _1(z_2, z-z_2) \rangle Y^V_1Y^\mu_1
\]
\[
=\int dz_1 \int dz_2
\langle
(\ki ^V (z_1)+(z_2-z)\ko ^V (z_1))
 (\kim (z_2) + (z_1-z)\kom (z_2))
\ran
Y^V_1Y^\mu_1
\]

\[=
[\int dz_1  
S_{1,1}^{\mu V}(\ko (z_1))+
\int dz_2  (z_2-z)A_1^\mu (\ko (z_2))]Y^\mu_1Y^V_1
\]

Consider the variations of both sides.

\[
 \delta \lan K_1^V K_1^\mu \ran Y^V_1Y^\mu_1
\]
\[
= \int dz' \int dz_1 \int dz_2 \lan \la (z') (\ko ^V (z_1) \kim (z_2) + 
\ki ^V (z_1) \kom (z_2) + 2\ko ^V (z_1) \kom (z_2) (z_2-z))\ran 
Y^V_1Y^\mu_1
\]
\[=
[\int dz' (\Lambda _{1,1}^\mu +\Lambda _{1,1}^V(z') \kom  
+2\Lambda _1(z')(z'-z)\kom )+
\]
\[
\int dz' \int dz_1 \Lambda _1(z')A_1^\mu (z_1)\underbrace {[\ko ^V (z')+
\ko ^V (z_1)]}_{1}
Y^V_1Y^\mu_1
\]

We also have
\[
\int dz_2 \delta A_1^\mu (\ko )(z_2-z) = \int dz_2 \Lambda _1\kom (z_2-z)
\]

Thus,
comparing as before, the gauge variations of both sides, we find
\be  \label{SVmu}
\int dz_2 \delta  S_{1,1}^{V\mu}(\ko ) = \int dz_2[ \Lambda _{1,1}^\mu (\ko )+ 
\Lambda_{1,1}^V (\ko )\kom + \Lambda _1 A_1 ^\mu + \Lambda _1\kom (z_2-z)]
\ee 

Finally, consider
\[
\lan K_2^V (z_1,z-z_1)\ran Y_2 ^V = S_2^V Y_2^V
\]
Varying we get
\[
\delta \lan K_2^V Y_2^V \ran =
\int dz' \lan [ \la _1(z')[\ki ^V (z_1) + (z_1-z)\ko ^V(z_1)]+
\la _2 (z')\ko ^V (z_1)\ran Y_2^V
\]
\[
=
\int dz_1 [\Lambda _{1,1}^V(z_1) + (z_1-z)\Lambda _1(z_1) + 
\Lambda _2(z_1)]Y_2^V
\]

Comparing gauge variations on both sides gives:
\be \label{S2V}
\int dz_1 \delta S_2^V = \int dz_1 [\Lambda _{1,1}^V + 
\Lambda _2 + (z_1-z)\Lambda _1]
\ee

At this point one can make the following observation: If we set
\[
\int dz' [\Lambda _{1,1}^V + (z'-z)\Lambda _1 ]
=\int dz' \Lambda _2
\]
we see that 
\[
\delta S_{1,1}^{V\mu}=\delta S_2^\mu
\]
\[
\delta S_{1,1}^{VV}=\delta S_2^V
\]

Thus as discussed in \cite{SZ,BSLV} 
for the free theory, we can reduce the number of degrees
of freedom in this theory to match that of critical string theory by setting
$\lan \ki ^V \ran = A_1^V=0$,  $S_{1,1}^{VV}=S_2^V$ and 
$S_{1,1}^{V\mu}=S_2^\mu$ consistently.

These identifications follow from the identification at the loop 
variable level:
\[
 K_1(z_1,z-z_1)^V\la _1(z_2) = \la _2 (z_1)
\]
and
\[
\ki ^V \kim = \ktm 
\]
and
\[
\ki ^V \ki ^V =  \kt ^V
\]
This is very similar to what was done at the free level in \cite{BSLV}.

We go back
to the loop variable equation for $Y_2^\mu$ from \cite{BSLV1} 
((5.3.39),(5.3.40), (5.3.45), 
(5.3.50) , (5.3.51) and (5.3.54)) 
We consider the terms that involve a contraction of the `$V$' index.
 These are
\[
\lan - e^{\int \int \ko (z_3).\ko (z_4) \tG}
\int dz_1 \int dz_2 \ki (z_1).\ko (z_2)i\int dz \kim Y_2^\mu \e \ran
\]
\[ 
=\int
[-iA_1^V(p)A_1^\mu (q)(z_1-z_2)^{2p.q} e^{i(p+q)Y}\eps ^{(p+q)^2}
-iS_{1,1}^{V\mu}e^{ikY}\eps^{k^2}]Y_2^\mu 
\]
\be   \label{i}
=-iS_{1,1}^{V\mu}\int e^{ikY}\eps^{k^2}Y_2^\mu 
\ee
where we have set $A_1^V=0$.

\[
 \lan - e^{\int \int \ko (z_3).\ko (z_4) \tG}
\int dz_1 \int dz_2 k_2 (z_1)^v\ko ^V(z_2) i\kom \int Y_2^\mu \e \ran
\]

\be   \label{ii}
=-S_2^Vi\kom \eps ^{k^2}\int e^{ikY}Y_2^\mu
\ee

\[
 \lan - e^{\int \int \ko (z_3).\ko (z_4) \tG}
\int dz_1 \int dz_2 \ki (z_1)^V\ki ^V(z_2) i\kom \int Y_2^\mu \e \ran
\]
\[
=\int
[\eps^{k^2}S_{1,1}^{VV}(k)\i\kom +
\eps^{(p+q)^2}A_1^V(p)A_1^V(q)i(p+q)^\mu e^{i(p+q)Y}(z-w)^{2p.q}]Y_2^\mu
\]
\be  \label{iii}
=\eps^{k^2}S_{1,1}^{VV}(k)i\kom \int Y_2^\mu e^{ikY} 
\ee

\[
 \lan - e^{\int \int \ko (z_3).\ko (z_4) \tG}\ko ^V \ko ^V i\ktm \int 
\e Y_2^\mu 
\ran
\]
\be   \label{iv}
=
\eps^{k^2}iS_2^\mu \int e^{i\ko Y}Y_2^\mu
\ee

When we apply the variations (\ref{AV}),(\ref{SVV}),(\ref{S2V}),(\ref{SVmu})
to (\ref{i}),(\ref{ii}),(\ref{iii}) and (\ref{iv}), we find that they add up
to zero. \footnote{Note also that they add upto zero, 
(to leading order in $z-w$)
even if we don't work with the reduced set of fields, i.e. even without
setting $A_1^V=0$ or making any of the other identifications,
the equations are gauge invariant. One can speculate that this describes some 
version of a non critical string theory.} Thus the full (tree level)
equations of string theory for the vertex operator $Y_2^\mu e^{ikY}$ are the
sum of the terms in the last section (with indices running 26 dimensions)
plus the terms calculated in this section.

This illustrates the dimensional reduction and compatibility with
gauge invariance. 
As explained earlier the contribution
of all the infinite number of fields have to be 
included in order for the cancellation to be exact.

\section{Uniform electromagnetic field: Born-Infeld Action}

We discuss, in this section, the loop variable approach in a 
non-perturbative setting. This is the situation of a uniform electromagnetic
field. In the sigma model approach this can be done to all orders in the
 field strength. This gives the Born-Infeld action. More precisely, it gives 
the $\beta$ -function, which, as emphasised in \cite{P,CANY,BSPT,BSZ}, is only 
{\em proportional} to the equation of motion. The proportionality constant
is the Zamolodchikov metric \cite{Z}.
We will show in this section how this is done in the loop 
variable approach. As in the perturbative case it involves writing down a
suitable wave functional. The main purpose of this section is to reproduce,
using loop variables,
the results in the literature that have been obtained using the sigma model
or more precisely boundary conformal field theory techniques. Gauge invariance
is not an issue here since only the field strength enters the equation
of motion.

The loop variable approach involves writing down (we  set $\xn =0$ since we are
not concerned with gauge invariance) the following object:
\[
\int \int {\cal D}\kom (t){\cal D}\kim (t)
e^{-i\int dt [\kom (t)X^\mu (t) +\kim (t) \p _tX^\mu (t)]}
\]
\be \label{1}
e^{{1\over 2}\int \int dtdt'
[\sum _{n,m=0,1}\km ^\mu (t) {\cal G}_{m,n}^{\mu \nu}(t,t')\kn ^\nu(t')]}
\Psi [\ko (t),\ki (t)].
\ee
Here ${\cal G}$ is the matrix
\be
\left(
\begin{array}{cccc}
 G & 0 & \p _{t'}G & 0 \\
0&G&0&\p _{t'}G\\
\p _tG&0&\p _t \p _{t'}G & 0\\
0&\p _t G&0&\p _t \p _{t'}G 
\end{array} \right)
\ee
 where the row vector, $k_m ^\mu$, ($\mu$ is the Lorentz index and $m$ is the
 mode index)  
 multiplying the matrix, is written 
 in the 
following order:
 $(\ko ^0 \, \ko ^1 \,
\ki ^0 \, \ki ^1)$.  $G(t,t') =\langle X(t) X(t')\rangle$ where
$t,t'$ are points on the boundary of the string world sheet. This could be a
disk or the upper half plane. $G$ satisfies the identity 
$\int dt'' \p _tG(t,t'')\p _{t''} G(t'',t')=1$ \cite{CANY,FT2}.

The wave functional in this case is
\be  \label{1.03}
\Psi [\ko (t),\ki (t)]=
\int {\cal D}X(t) e^{-i\int dt [\kom (t)X^\mu (t)]} 
\prod _{\mu t}
\delta [\kim (t) - A^\mu (X(t)]
\ee

where $A^\mu = 1/2F^{\mu \nu}X^\nu$.
For simplicity we take the two dimensional case, where $\mu =0,1$.
Thus the delta function becomes
\[
\prod _t \delta [X^1(t)-{2\ki ^0(t)\over F}]
\delta [X^0(t)+ {2\ko ^1(t)\over F}]{1\over 
F^2}
\]

Here $F=F^{01}=-F^{10}$.

Doing the the integral gives
\[
\Psi [\ko (t),\ki (t)]=
e^{{2i\over F}\int dt [\ko ^0(t)\ki ^1(t) - \ko ^1(t)\ki ^0(t)]}
\prod _t{1\over F^2}
\]

When we substitute this expression in (\ref{1}), we get

\be \label{PF}
\int {\cal D}\ko (t){\cal D}\ki (t)
e^{{1\over 2}\int \int dtdt'
[\sum _{n,m=0,1}\km ^\mu (t) {\cal G}_{F,m,n}^{\mu \nu}(t,t')\kn ^\nu(t')]}
e^{-i\int dt [\kom (t)X^\mu (t) +\kim (t) \p _tX^\mu (t)]}
\ee
where 
${\cal G}_F$ is the matrix
\[
\left(
\begin{array}{cccc}
 G & 0 & \p _{t'}G & {2i\over F} \\
0&G&-{2i\over F}&\p _{t'}G\\
\p _tG&-{2i\over F}&\p _t \p _{t'}G & 0\\
{2i\over F}&\p _t G&0&\p _t \p _{t'}G 
\end{array} \right)
\]

The Gaussian integral can easily be done and using the identity obeyed by $G$
we get 
\[
e^{{1\over 2}\int \int dt dt' {\bf X}^T {\cal G}_F^{-1}{\bf X}}
Det^{-1/2} [F^{-2}(1+F^{-2})\delta (t-t')]=
\]
\[
e^{{1\over 2}\int \int dt dt' {\bf X}^T {\cal G}_F^{-1}{\bf X}}
 Det[F^2(1+F^2)^{-1/2}\delta (t-t')]
\]

Here ${\bf X}= (X^0 ,\,  X^1,\,  \p _t X ^0, \, \p _t X ^1 )$

Combining the $1\over F^2$ in the wave functional we get
\[
e^{{1\over 2}\int \int dt dt' {\bf X}^T {\cal G}_F^{-1}{\bf X}}
Det [1+F^2]^{-1/2}\delta (t-t')
\]
Using zeta function regularization (as explained in \cite{FT2}) the 
determinant gives
$[1+F^2]^{+1/2}$.
Thus the final answer is 
\be   \label{ZX}
Z[{\bf X}]= 
e^{{1\over 2}\int \int dt dt' {\bf X}^T {\cal G}_F^{-1}{\bf X}}
[1+F^2]^{+1/2}.
\ee

 $Z[0]$ is the Born Infeld Lagrangian. The precise connection 
between the sigma model ``partition function'' and the actual 
space-time effective action of the   fields of the string
is not clear to us, although in this particular example they seem to coincide.
What we do know is that the $\beta$-function gives the equation of motion
upto the Zamolodchikov metric prefactor. \footnote{It 
was also shown in \cite{BSZ}
that the proper-time equation gives the full equation.}
The equations obtained in the loop variable approach are the equivalent of the 
$\beta$-function. They are a gauge invariant generalization obtained by
making the cutoff depend on the world sheet location.

One can obtain the equation for the photon by looking at the 
coefficient of $\p _t X^\mu$ in the conformal variation of the above
``generating functional with sources X'', (\ref{ZX}). In the loop
variable approach this was done by replacing $G$ by  $G+\Sigma$ 
and the
coefficient of $\Sigma$ gave the equation of motion.\footnote{We remind the
reader that all the $\xn$ variables used in the loop variable approach have
been set to zero, as we are not concerned with gauge invariance 
related issues.} It is clear here that there are no terms linear in $X$
and the equation is trivial. In order to get a non-trivial equation one has 
to include a perturbation describing the non-uniformity of $F$.

Thus we change the expression for $A$ into:
\be  \label{1.05}
A^\mu (X) = {1\over 2}F^{\mu \nu}X^\nu + 
{1\over 3} \p _\rho F^{\mu \nu}X^\rho X^\nu
\ee

The delta function in (\ref{1.03}) changes to
\[
\prod _t \delta [\kim (t) - {1\over 2}F^{\mu \nu}X^\nu (t) + 
{1\over 3} \p _\rho F^{\mu \nu}X^\rho (t) X^\nu (t)]
\]
\[
=(1+\int dt'{1\over 3} \p _\rho F^{\mu \nu}X^\rho(t') X^\nu (t')]
{\delta \over \delta \kim (t')} )
\prod _t \delta [\kim (t) - {1\over 2}F^{\mu \nu}X^\nu (t)] 
\] 
 
If we insert this into (\ref{1.03}) we get
for the modified wave functional,

\[
\Psi [\ko (t),\ki (t)]=
\int {\cal D}X(t) e^{-i\int dt [\kom (t)X^\mu (t)]} 
\]
\[
(1+ \int dt'{1\over 3} \p _\rho F^{\mu \nu}X^\rho(t') X^\nu (t')]
{\delta \over \delta \kim (t')})
\prod _{\mu t}
\delta [\kim (t) - A^\mu (X(t)]
\]

\be   \label{1.07}
=
(1- \int dt'{1\over 3} \p _\rho F^{\mu \nu}i{\delta \over \delta k_0^\rho (t')}
i{\delta \over \delta \kon (t')}){\delta \over \delta \kim (t')})
\Psi [\ko (t),\ki (t)]
\ee

We insert the modified wave functional into (\ref{1}), integrate by parts
on $\ki ,\ko$ and pick the piece proportional to $\dot X$:

\[
\int \int {\cal D}\kom (t){\cal D}\kim (t)
\int dt' {1\over 3}\p _\rho F^{\mu \nu}i\p _{t'} X^\mu (t')
[ {\cal G} ^{0\rho , 0\nu}(t',t') + ({\cal G}k)^{0\rho}({\cal G}k)^{0\nu}]
\]
\[
e^{-i\int dt [\kom (t)X^\mu (t) +\kim (t) \p _tX^\mu (t)]}
\]
\be 
e^{{1\over 2}\int \int dtdt'
[\sum _{n,m=0,1}\km ^\mu (t) {\cal G}_{m,n}^{\mu \nu}(t,t')\kn ^\nu(t')]}
\Psi [\ko (t),\ki (t)].
\ee 

Replacing $\kom$ by $i{\delta \over \delta X^\mu (t')}$ we get
and
$\kim$ by $i{\delta \over \delta \dot X^\mu (t')}$,
(We let $X^\alpha$ represent $X^\mu, \dot X^\mu$)
\[
\int dt' {1\over 3}\p _\rho F^{\mu \nu}i\p _{t'} X^\mu (t')
\]
\[
[ {\cal G} ^{0\rho , 0\nu}(t',t') + \int dt'' \int dt'''
({\cal G}(t',t'')^{0\rho,\al}i{\delta \over \delta X^\al (t'')}
({\cal G}^{0\nu,\beta}(t',t''')i{\delta \over \delta X^\beta (t''')}]
\]
\[
e^{{1\over 2}\int \int dt dt' {\bf X}^T {\cal G}_F^{-1}{\bf X}}
[1+F^2]^{+1/2}.
\]
\[
=
\int dt' {1\over 3}\p _\rho F^{\mu \nu}i\p _{t'} X^\mu (t')
\]
\[
[{\cal G}^{0\rho ,0\nu}(t',t')-({\cal G}{\cal G}_F^{-1}{\cal G})^{0\rho ,0\nu}(t',t')]Z[{\bf X}]
\]

One can easily evaluate the expression multiplying $Z[{\bf X}]$. It is
 the matrix
\be
\left( \begin{array}{cc}
{G(t,t')\over 1+F^2} & 
\int dt''iFG(t,t'')\p _{t''}G(t'',t')\\
 -\int dt''iFG(t,t'')\p _{t''}G(t'',t')
&{G(t,t')\over 1+F^2} \end{array} \right)
\ee
multiplied by $\delta (t'-t)$.

The off diagonal elements are antisymmetric in $t,t'$ and therefore
vanish when $t=t'$. Thus we get
\[
 \delta ^{\rho \nu}\int dt {G(t,t')\over 1+F^2}\delta (t-t')  
\]

When we make a conformal transformation the change in $G$ is $\Sigma$ so we get
a term
\be \label{1.10}
{1\over 3}\p _\rho F^{\mu \nu}i\p _{t'} X^\mu (t')
\delta ^{\rho \nu}\int dt {\Sigma (t,t')\over 1+F^2}\delta (t-t')  
={1\over 3}\p _\rho F^{\mu \nu}i\p _{t'} X^\mu (t')
 \delta ^{\rho \nu} {\Sigma (t',t')\over 1+F^2}
\ee
The coefficient of $\Sigma $ in the above expression  is one contribution to 
the $\beta$-function. Variation of
$\delta (t-t')$ gives terms proportional to delta function and it's
 derivatives (times the parameter of conformal transformation). Multiplying
$G$ this is singular and introduces further powers of the cutoff. These are
equivalent to $(ln \, a)^2$ divergences and do not contribute to the 
$\beta$-function. Thus  the coefficient of $\Sigma$ in 
(\ref{1.10}) is the full expression for the $\beta$-function.

As explained in \cite{BSZ} one can get the full equation of motion 
(= $\beta$-function $\times$ Zamolodchikov metric) if one uses
the proper time equation. But the method here gives only the $\beta$-function.
That this answer is correct can be seen by consulting \cite{CANY}.

\section{Conclusions}

In this paper we have given some examples of loop variable calculations.
We have tried to make concrete, 
some of the ideas described in \cite{BSLV1,BSLV2}. We have given examples
of the wave functional that is ubiquitous in these two papers. We have also
worked out in detail the equations of a massive mode 
interacting with electromagnetism.  Most importantly while a formal argument
of the gauge invariance was given in \cite{BSLV2}, there was not a detailed
understanding of how the infinite tower of massive modes contribute 
in the gauge invariance of any one equation. 
This understanding 
has now been obtained, as described in Section 3.

Before dimensional reduction, all the modes are massless and we are not
describing critical string theory - at least not in any recognizable way.
Thus dimensional reduction is crucial for making contact with string theory.
This was described in a general way in \cite{BSLV2}. Here we have worked
out all the details in this particular example. We have also shown how the
truncation of fields necessary for making contact with string theory works.

Finally, by reproducing the Born -Infeld equations,
we make contact with the open string sigma model 
(or boundary conformal field theory) approach. This was useful in explaining
the ideas of \cite{BSLV1,BSLV2} in more conventional terms. 

It is tempting to speculate that the higher dimensional massless theory
is some more ``symmetric'' phase of string theory. The idea that the 
interactions emerge naturally by broadening the string to a band is reminiscent
of membranes. All this points to M-theory. But we do not see any way of making
a connection. 

As another interesting test one could try to reproduce the results
on the tachyon that have been obtained in the literature using other 
 techniques such as String Field Theory or Background Independent
String Field Theory \cite{KS,AS2,ASZ,GS,W2,WL,SS}.

\end{document}